\documentclass[twocolumn,10pt]{asme2ej}

\usepackage{graphicx} 
\usepackage{xcolor} 
\usepackage{amsmath,amssymb,accents,color}
\title{Technical Brief: Finite Element Modeling of Tight Elastic Knots}

\author{Changyeob Baek
    \affiliation{
	Department of Mechanical Engineering\\
	Massachusetts Institute of Technology\\
	Cambridge, MA 02139\\
    Email: cbaek@mit.edu
    }	
}

\author{Paul Johanns
    \affiliation{Flexible Structures Laboratory\\
	Institute of Mechanical Engineering\\
	\'{E}cole Polytechnique F\'{e}d\'{e}rale de Lausanne\\
	Lausanne 1015, Switzerland\\
    Email: paul.johanns@epfl.ch
    }
}

\author{Tomohiko G. Sano
    \affiliation{Flexible Structures Laboratory\\
	Institute of Mechanical Engineering\\
	\'{E}cole Polytechnique F\'{e}d\'{e}rale de Lausanne\\
	Lausanne 1015, Switzerland\\
    Email: tomohiko.sano@epfl.ch
    }
}

\author{Paul Grandgeorge
    \affiliation{Flexible Structures Laboratory\\
	Institute of Mechanical Engineering\\
	\'{E}cole Polytechnique F\'{e}d\'{e}rale de Lausanne\\
	Lausanne 1015, Switzerland\\
    Email: paul.grandgeorge@epfl.ch
    }
}

\author{Pedro M. Reis\thanks{To whom correspondence should be addressed: pedro.reis@epfl.ch} \\
    \affiliation{Flexible Structures Laboratory\\
	Department of Mechanical Engineering\\
	Institute of Mechanical Engineering\\
	\'{E}cole Polytechnique F\'{e}d\'{e}rale de Lausanne\\
	Lausanne 1015, Switzerland\\
    Email: pedro.reis@epfl.ch
    }
}

\usepackage{color}

\begin{document}

\maketitle    

\begin{abstract}
{\it We present a methodology to simulate the mechanics of knots in elastic rods using geometrically nonlinear, full three-dimensional (3D) finite element analysis. We focus on the mechanical behavior of knots in tight configurations, for which the full 3D deformation must be taken into account. To set up the topology of our knotted structures, we apply a sequence of prescribed displacement steps to the centerline of an initially straight rod that is meshed with 3D solid elements. Self-contact is enforced with a normal penalty force combined with Coulomb friction. As test cases, we investigate both overhand and figure-of-eight knots. Our simulations are validated with precision model experiments, combining rod fabrication and X-ray tomography. Even if the focus is given to the methods, our results reveal that 3D deformation of tight elastic knots is central to their mechanical response. These findings contrast to a previous analysis of loose knots, for which 1D centerline-based rod theories sufficed for a predictive understanding. Our method serves as a robust framework to access complex mechanical behavior of tightly knotted structures that are not readily available through experiments nor existing reduced-order theories.
}
\end{abstract}

\section{Introduction}
Knots are ubiquitous mechanical links used to establish kinematic constraints between filaments via friction. Knotted filaments are utilized regularly in shoelaces~\cite{daily2017roles}, sailing and climbing ropes, and surgical threads~\cite{zimmer1991influence, uehara2007effects, philoadvanced}. Accordingly, there have been numerous efforts to quantify the mechanical performance of knots with various topologies, through experiments~\cite{pieranski2001localization, uehara2007effects, Patil2020}, numerical simulations~\cite{konyukhov2010geometrically, Durville2012, qwam2018validation, Patil2020}, and modeling based on geometric strings (\textit{i.e.}, neglecting elasticity)~\cite{Maddocks1987a, katritch1996geometry, grosberg1996flory, gonzalez1999global, pieranski2001localization} and elastic rods~\cite{Audoly2007, clauvelin2009matched, Jawed2015untangling}. The one-dimensional (1D) elastic rod models derived in~\cite{Audoly2007, clauvelin2009matched, Jawed2015untangling} were based on the theory of elastic Kirchhoff rods, with a centerline description. The fundamental basis of this rod theory is the assumption of negligible cross-sectional deformations and the inextensibility of the centerline of the rod. These simplifying assumptions allow for an understanding of elastic knots, especially in their \textit{loose} configurations. However, due to the complex interplay between various modes of three-dimensional deformation that emerge in functional knots, Kirchhoff-like rod models cannot capture the mechanical behavior of \textit{tight} elastic knots.

Here, we present a computational framework that allows for the systematic investigation of tight elastic knots based on a fully 3D finite element method (FEM). We implement our simulation framework for the illustrative cases of overhand and figure-of-eight knots.
In parallel, the numerical results are validated against experiments using precise rod fabrication, mechanical testing, and tomographic imaging. We clarify the prominent role of 3D deformation in tight elastic knot configurations that we contrast with the existing theory of loose knots. Throughout, we place more emphasis on the method that we have employed, rather than on results, hoping that this technical brief will instigate further analyses in future research.

Our paper is organized as follows. First, in Sec.~\ref{sec:problem}, we identify the key parameters of overhand knots and provide an overview of the theory for the mechanical behavior of loose knots developed previously by Audoly, Clauvelin \& Neukirch~\cite{Audoly2007}. Next, in Sec.~\ref{sec:methods}, we focus on the numerical FEM procedure to tie a knot. In Sec.~\ref{sec:experiments}, we detail the experimental protocol to fabricate elastomeric rods and the tools that were developed to validate the numerical results. Further, we compare numerical and experimental results, such as force-displacement curves and curvature profiles along the knotted rod's centerline in Sec.~\ref{sec:results}. We also compare these numerical and experimental results to the theoretical prediction provided in~\cite{Audoly2007}. Finally, we provide some concluding comments in Sec.~\ref{sec:conclusion}. 

\section{Definition of the Problem}
\label{sec:problem}

\begin{figure}
    \centering
    \includegraphics[width=0.8\columnwidth]{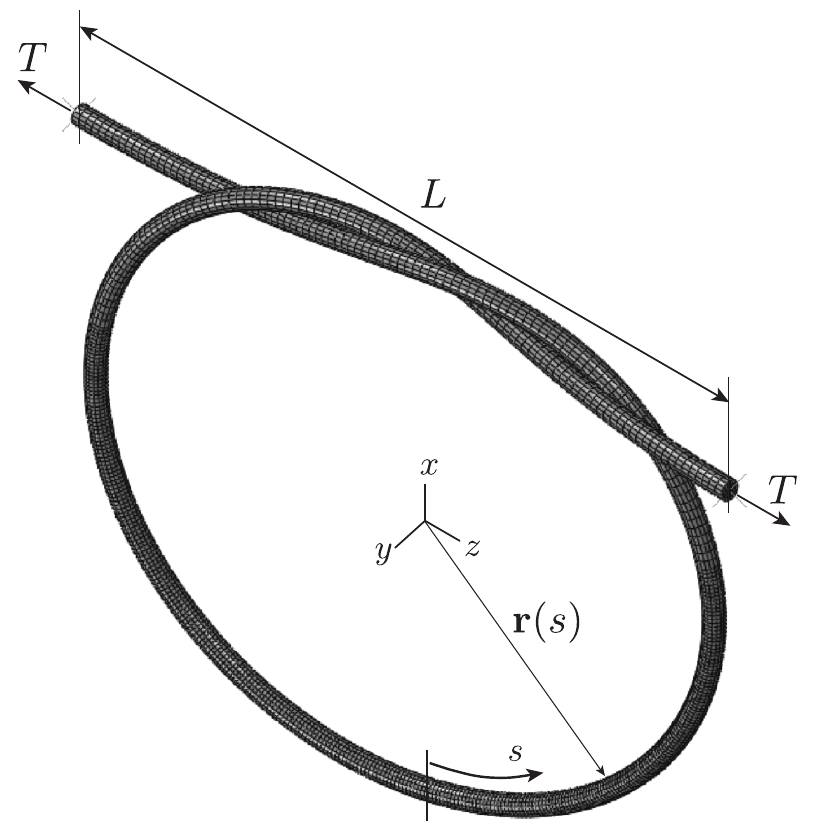}
    \caption{An elastic rod of undeformed length, $L_0$, and undeformed diameter, $D$, is knotted into a simple overhand knot configuration. The end-to-end distance, $L$, is regarded as the control variable. The corresponding tensile force, $T$, and the position of the material centerline of the knot, $\mathbf{r}(s)$, are measured.}
    \label{fig:problemdef}
\end{figure}

In Fig.~\ref{fig:problemdef}, we present a representative snapshot of an \textit{elastic overhand knot}, the central topic of this technical brief. The tying of a figure-of-eight knot, which we have also addressed, will be discussed in Sec.~\ref{sec:results}. An elastic rod of Young's modulus $E$, undeformed length $L_0$, undeformed diameter $D$, and dynamic Coulomb friction coefficient $\mu$ for self-contact is deformed into a simple overhand knot configuration. The centerline position of the elastic knot is denoted as $\mathbf{r}(s)$, where $s$ is the arc-length coordinate of the centerline of the rod. The two extremities of the rod are aligned along the $z$-axis. While one extremity is clamped, the other is free to twist about the $z$-axis. The end-to-end distance between the extremities is $L$, and the corresponding tensile force required for tightening is $T$. Following the work of Jawed \textit{et al.}~\cite{Jawed2015untangling}, we define the normalized \textit{end-to-end shortening}, $\bar{e}=e/D=(L_0-L)/D$, as the single control variable of this problem. The value of $\bar{e}$ decreases as the knot is tightened. We will be interested in quantifying the centerline position, $\mathbf{r}(s;\,\bar{e})$, and the applied tension, $T(\bar{e})$, as functions of $\bar{e}$.

For completeness of our discussion, we briefly review an existing model for the mechanical response of \textit{loose elastic overhand knots} based on the theory of Kirchhoff rods. According to the previous work by Audoly, Clauvelin, and Neukirch~\cite{Audoly2007}, the tensile force, $T$, of a loose overhand knot can be expressed as
\begin{equation}
    \frac{T{D}^2}{4B} = \frac{\pi^2}{2 \bar{e}^2} \pm \frac{0.492 \mu \pi^{1.5}}{\bar{e}^{1.5}},
    \label{eq:knot_audoly}
\end{equation}
where $B=\pi E D^4/64$ is the bending stiffness of the rod. The first term in Eq.~\eqref{eq:knot_audoly} represents the elastic bending contribution, and the second term arises due to dynamic friction in the regions of self-contact. The positive (or negative) sign of the second term corresponds to the tying (or untying) of the knot when $\dot{\bar{e}}<0$ (or when $\dot{\bar{e}}>0$). 
Note that, in the limit of $\bar{e}\rightarrow 0$, Eq.~\eqref{eq:knot_audoly} predicts the divergence of $T$, which is a manifestation of the inextensibility constraint of the underlying Kirchhoff rod theory. Ref.~\cite{Audoly2007} reported that Eq.~\eqref{eq:knot_audoly} is valid for loose-knot, in the $\bar{e} \gtrsim 500$ regime. 

In the present study, we seek to extend the scope of past studies on the mechanics of knots by focusing on tight configurations. Specifically, we address the mechanics of \textit{tight} overhand knots, in the $\bar{e} \lesssim 100$ regime, as well as tight figure-of-eight knots. Furthermore, our study will consider situations with moderate friction ($\mu = 0.32$), whereas Ref.~\cite{Audoly2007} was limited to weak friction ($\mu \simeq 0.1$). The mechanical response of knots in tight configurations is significantly more complex than that for loose knots. Hence, instead of following the reduced-order analytical methods of past studies, we develop a methodology to simulate knotted structures in filaments. For this purpose, we employ the framework of fully three-dimensional (3D) finite element analyses.

\section{Finite Element Procedure for Tying Elastic Knots}
\label{sec:methods}

Our simulation approach is based on the finite element method (FEM) using the commercially available software package \texttt{ABAQUS/STANDARD}. We conducted a nonlinear dynamic-implicit analysis to capture the geometrically nonlinear deformation of right knots tied into an elastic rod. An initially straight elastic rod was meshed with reduced hybrid 3D solid elements (\texttt{C3D8RH}). The level of discretization of the mesh on the cross-section of the rod was $\sim 100$. Along the axial direction, the level of discretization was varied depending on the aspect ratio of the rod, to ensure that the elements maintained a regular cubic shape. 

Even if we could have incorporated different material models in our FEM simulations, we focused exclusively on elastic materials. The rationale for this choice is twofold. First, the experimental results used to validate the simulations (presented in Sec.s~\ref{sec:experiments} and~\ref{sec:results}) were obtained using rods made out of vinylpolysiloxane (Young's modulus, $E=1.25\,\text{MPa}$; VPS-32, Zhermack). VPS is an elastomeric material that we modeled as a neo-Hookean incompressible solid. The goal was to have a direct map between the simulations and experiments for quantitative validation. Secondly, the focus on the elastic case allowed us to emphasize the high-fidelity of the simulations in terms of the topological preparation protocol and the appropriateness of the frictional contact interactions, without further complexifying the problem with additional constitutive ingredients.

Self-contact frictional interactions in the rod were taken into account by enforcing a normal penalty force combined with a tangential frictional force, with a prescribed dynamic Coulomb friction coefficient, $\mu=\{0,0.32\}$. Note that we studied both frictionless and frictional elastic knots.
\begin{figure}
    \centering
    \includegraphics[width=\columnwidth]{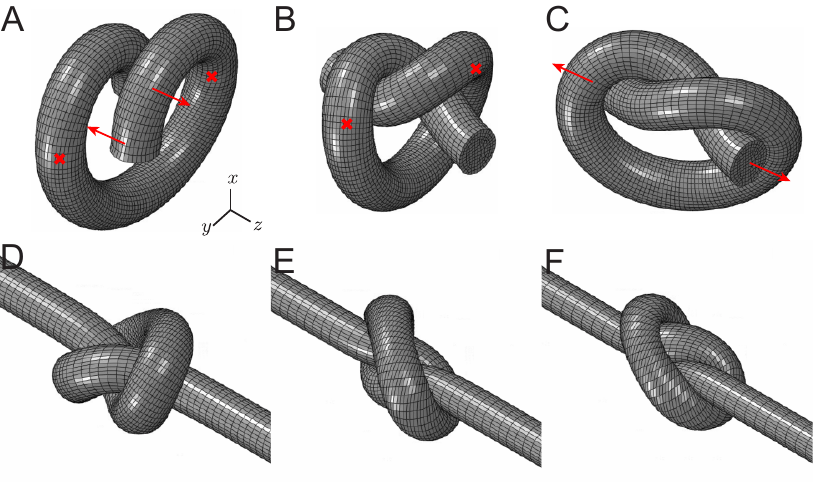}
    \caption{Numerical protocol and pathing procedure to tie an overhand knot. (A) An originally straight rod is first bent to fix a pair of intermediary points (x-shaped symbols). Then, its extremities are pulled (red arrows) along the $\pm z$ directions. (B) As a result, the topology of the knotted configuration is established. (C) We then remove the positional constraints, formerly the x-shaped symbols in (A) and (B), to obtain an equilibrium configuration. (D)-(F) Having established the topology of the knot, we then tighten it by controlling the positions of the extremities to set a given value of the end-to-end shortening, $e$.}
    \label{fig:sequence}
\end{figure}

Throughout the simulations, the extremities of the rod are kinematically tied to a pair of control nodes located at each of the ends. The topology of the knotted rod was established by applying a sequence of prescribed displacements and rotations to these control nodes. We adopted a loading sequence from previous work on reduced FE modeling of knots~\cite{Durville2012}. The sequence of simulation snapshots in Fig.~\ref{fig:sequence} illustrates the tying process of an overhand knot, involving the following four steps: 

\begin{enumerate}
    \item[(i)]  Firstly, we bent the rod into the configuration depicted in Fig.~\ref{fig:sequence}A, while fixing the position of a pair of auxiliary nodes (denoted as x-shaped symbols in the figure) on the material centerline of the rod. At this stage, the extremities of the rod faced the $-x$-direction.
    \item[(ii)] Secondly, the extremities of the rod were displaced inside the loop (see Fig.~\ref{fig:sequence}B), now facing the $\pm z$-direction, thereby establishing the knotted configuration. During this second step, the positional constraints applied to the auxiliary nodes were still enforced so that those points lay on the $xy$-plane. 
    \item[(iii)] Thirdly, we removed the imposed constraints on the auxiliary points to allow for the equilibrium configuration of the knot to be achieved (see Fig.~\ref{fig:sequence}C). 
    \item[(iv)] Finally, we tightened the knot (see Fig.~\ref{fig:sequence}D to E) by continuously decreasing the normalized end-to-end shortening, $\bar{e}$, at the constant speed, $\dot{e}=-0.5\,\mathrm{mm/s}$, and measured the quantities of interest, including the tensile force, $T(\bar{e})$, and the configuration of the knot, $\mathbf{r}(s;\,\bar{e})$.
\end{enumerate}

The typical computational cost for a full knot-tying simulations of an overhand knot, in which the normalized end-to-end shortening decreased gradually from $\bar{e}=20$ to $\bar{e}=-10$ (in steps of $\Delta\bar{e}=0.15$), is approximately 60 hours on a desktop workstation with an octa-core processor (Intel Xeon processor 6136 3.20\,GHz) and 32\,GB of RAM.

\section{Experimental Procedure}
\label{sec:experiments}

In Sec.~\ref{sec:results}, we will validate the numerical protocol introduced above against precision experiments, the details of which are provided next. To obtain the precision experimental data used for validation, we  (\textit{i}) measured the macroscopic mechanical response through the relation between the tensile force $T$ and end-to-end shortening $e$, and (\textit{ii}) thoroughly quantified coordinates of the centerline of the knotted polymeric rod. The elastic rods used for either the experiments in (\textit{i}) or (\textit{ii}) required slightly different fabrication protocols, we detail next.

For the experiments in (\textit{i}) -- mechanical characterization of $T$ \textit{vs.} $e$ --, we cast our rod out of the vinylpolysiloxane (VPS32, Elite Double 32, Zhermack, Young's modulus $E=1.25$~MPa, density $\rho=1.160$~kg/m$^3$), inside a straight steel cylinder (SS pipes, part number PSTS12A-400, Misumi). This cylindrical mold had an inner diameter of $D=8.3$\,mm, and length~400\,mm. Upon curing of the VPS32 (in $\sim$30\,mins), we unmolded the solid elastomeric rod and let it rest for seven days to ensure the steady-state of its mechanical properties. After this resting period, we cut our undeformed (straight) elastomeric rod to the desired length, $L_0$. We then tied the appropriate knot (overhand or figure-of-eight) into the rod and performed tensile tests under clamped boundary conditions. We used a universal testing machine (Instron 5943, Norwood, MA) to obtain the $T(e)$ experimental curves. During testing, the bottom clamp was mounted onto a rotary air-bearing to ensure a twist-free boundary condition there and enable a direct comparison to Ref.~\cite{Audoly2007}.  In this mechanical characterization experiments, we explored both the \textit{frictionless} and \textit{frictional} cases. For the the \textit{frictionless case}, a few drops of silicone oil (Bluesil 47V1000, Slitech, dynamic viscosity~1\,Pa$\cdot$s) were applied to the contact regions of the knot. The viscosity of the silicone oil generated a thin lubrication layer that reduced local tangential forces significantly. For the experiments in the \textit{frictional case}, after curing of the VPS, the surface of the rod was conditioned with talcum powder (Milette baby powder, Migros, Switzerland) that was adsorbed by the VPS surface during the following 24~hours. Gently wiping off the excess talcum powder from the surface with a fine cloth ensured Amontons-Coulomb frictional behavior, with a robust dynamic friction coefficient, $\mu=0.32\pm0.03$. 

For the experiments in (\textit{ii}) -- quantification of the location of the physical centerlines of the knotted rod --, the rod fabrication protocol presented above had to be slightly modified to make them compatible with our imaging technique. We obtained volumetric images of the elastic knots using X-ray micro-computed tomography ($\mu$CT100, Scanco Medical, Switzerland). Measuring the location of the physical centerline of the elastomeric rods in the 3D images required us to introduce a thin concentric physical centerline fiber (diameter 500\,$\mu$m) inside the VPS32 rods. To fabricate this concentric centerline fiber, we added a nylon filament that was taut straight, concentrically to the steel cylinder (the mold), before casting. After curing of the VPS32, we pulled out the nylon filament, leaving a thin and hollow cylindrical void along the central axis. Next, we filled the void left by the removed nylon filament with a different silicone polymer (Solaris, Smooth-On; Young's modulus 340\,kPa, and density 1001\,kg/m$^3$). The density contrast between VPS32 and Solaris was sufficient to allow for differentiation and segmentation between the regions of these two polymers -- physical centerline (Solaris) and the bulk of the rod (VPS32) -- on the 3D $\mu$-CT images of the elastic knots. Finally, the 3D images of the elastic knots were post-processed to extract the precise locations of the physical centerline coordinates. We performed the centerline tracking using a Matlab code developed in-house that was based on an edge-detection algorithm. This code involved the iterative construction of orthogonal frames adapted to each point of the centerline. The automatic localization of the physical concentric Solaris fiber was based on the density contrast between Solaris and VPS32, using an edge-and-centroid detection algorithm.

\section{Results}
\label{sec:results}

\begin{figure}
    \centering
    \includegraphics[width=\columnwidth]{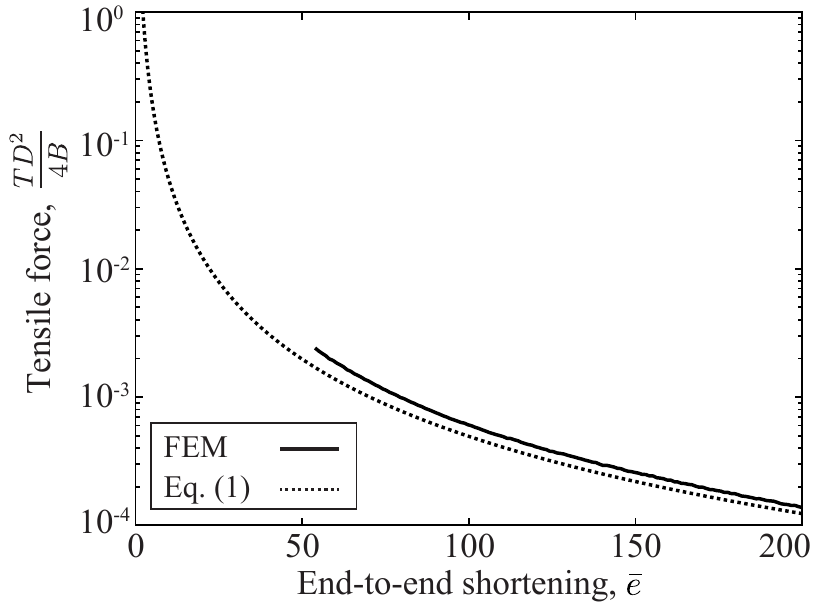}
    \caption{Normalized tensile force, $TD^2/(4B)$, versus the normalized end-to-end shortening, $\bar{e}$ for frictionless rod of length $L_0=800\,\text{mm}$, obtained from the FEM simulations (solid line). The data from the Kirchhoff rod model of an overhand knot (Ref.~\cite{Audoly2007}) is plotted as a dotted line.}
    \label{fig:tensileforce_overhand_loose}
\end{figure}

Having introduced the computational framework for the tying of tight knots in elastic knots, we first validate our FEM procedure by comparing the numerical results to the one-dimensional model of Ref.~\cite{Audoly2007}, for overhand knots in a \textit{loose} configuration. We consider an elastic rod with the following parameters: $L_0=2213\,\text{mm}$, $D=8.3\,\text{mm}$, and $\mu=0$. For this loose configuration, the end-to-end shortening is varied within the range $50 \le \bar{e} \le 200$, which falls into the borderline region between the loose-knot and the tight-knot regime, as defined in Ref.~\cite{Audoly2007}. In Fig.~\ref{fig:tensileforce_overhand_loose}, we plot the dimensionless tensile force, $T{D}^2/(4B)$, as a function of $\bar{e}$, obtained from the FEM simulations (solid line), onto which we juxtapose the prediction from Eq.~\eqref{eq:knot_audoly} (dotted line), with $\mu=0$. As $\bar{e}$ increases, we find that the FEM results approach the theoretical prediction Ref.~\cite{Audoly2007}, albeit with a slight offset (the FEM result is 11$\,\%$ higher than the theory at $\bar{e}=200$). We were not able to extend the simulations into the regime of validity of the theory ($\bar{e} \approx 500$), since this would have required an excessive computational power due to an unavoidable finer mesh. Still, we observe that the FEM simulation only starts to deviate from the theory more significantly in the tight-knot regime, $\bar{e} \lesssim 100$.

\begin{figure}
    \centering
    \includegraphics[width=\columnwidth]{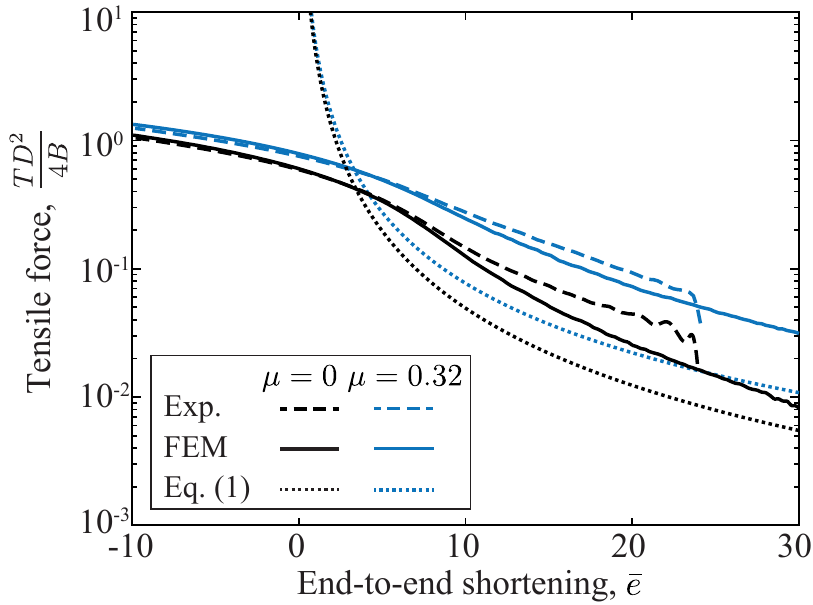}
    \caption{Normalized tensile force versus normalized end-to-end shortening for both frictionless ($\mu=0$) and frictional ($\mu=0.32$) elastic knots. The experiments (dashed lines) and FEM simulations (solid lines) were conducted for a rod with $L_0=350\,\text{mm}$ and $D=8.3\,\text{mm}$. The theoretical predictions (dotted lines) correspond to Eq.~\eqref{eq:knot_audoly} from Ref.~\cite{Audoly2007}. While the FEM simulations are in excellent agreement with the experiments, both deviate from the theoretical prediction in the regime of lower values of the end-to-end shortening.}
    \label{fig:tensileforce_overhand_tight}
\end{figure}

We now quantify the applied tension required to tie an overhand elastic knot in the tight-knot regime. Here, we consider a shorter rod (initial length, $L_0=350\,\text{mm}$) than the ones used above, in order to more effectively investigate the mechanical behavior of a tight overhand knot. The range of the values of end-to-end shortening considered in this tight regime is $-10 \le \bar{e} \le 30$, far below the threshold of validity of Eq.~\eqref{eq:knot_audoly} ($\bar{e} \sim 500$). In Fig.~\ref{fig:tensileforce_overhand_tight}, we plot the dimensionless tensile force, $\frac{T{D}^2}{4B}$, as a function of $\bar{e}$, from the precision experiments (dashed lines) and the FEM simulations (solid lines). The frictionless ($\mu=0$; black lines) and frictional cases for overhand knots ($\mu=0.32$; blue lines) are addressed and contrasted in both the experiments and simulations. The excellent agreement between the FEM results and the experimental data validates our numerical knot-tying procedure. By contrast, we observe a significant deviation between the numerical/experimental data and the prediction from Eq.~\eqref{eq:knot_audoly} for loose knots, regardless of whether friction is present or not. As mentioned in Sec.~\ref{sec:problem}, this discrepancy does not come as a surprise, given the significantly lower range of $\bar{e}$ that we have considered for our tight configurations compared to the regime of validity of the theory underlying Eq.~\eqref{eq:knot_audoly}.

\begin{figure}
    \centering
    \includegraphics[width=\columnwidth]{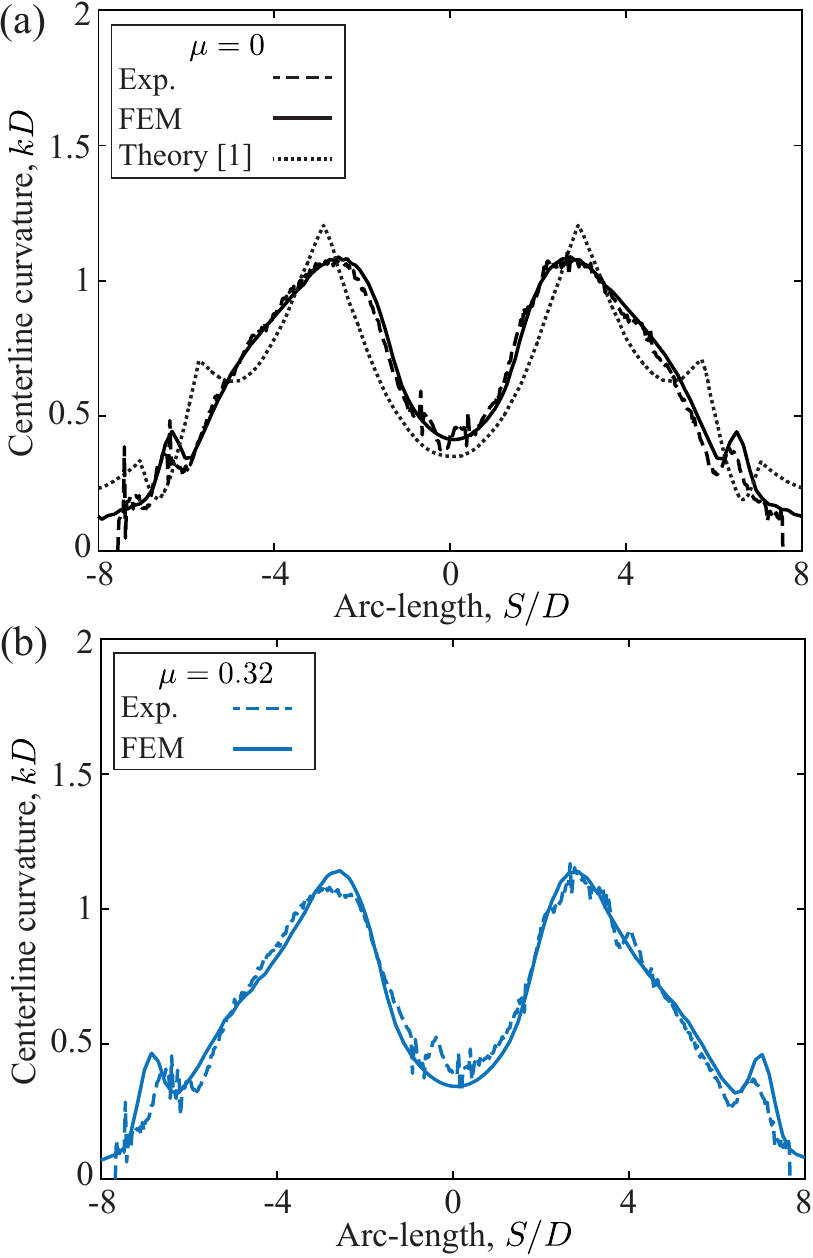}
    \caption{Profiles of the centerline curvature for an overhand knot with normalized end-to-end shortening, $\bar{e}=9.58$, obtained from the FEM simulations (solid lines), the experiments (dotted lines), and the theoretical prediction from Eq.~\eqref{eq:knot_audoly} (dashed line). The elastic rod onto which an overhand knot is tied has an initial length of $L_0=130\,\text{mm}$ and a self-contact friction coefficient of (a) $\mu=0$ and (b) $\mu=0.32$.}
    \label{fig:curvature}
\end{figure} 

Inspired by the success reported above in quantifying the tensile forces during knot-tying, we further quantify the shape of tight elastic overhand knots by contrasting the FEM simulations to precision X-ray micro-computed tomography ($\mu$-CT). A material centerline of the elastic knot, $\mathbf{r}(s)$, is digitized from the volumetric image acquired from the $\mu$-CT, as well as from the FEM simulations. Again, we considered both the frictionless ($\mu=0.0$) and frictional ($\mu=0.32$) cases. The initial length of the rod was $L_0=130\,\text{mm}$, the initial diameter was $D=8.3\,\text{mm}$, and the end-to-end shortening was $\bar{e}=9.58$. In Fig.~\ref{fig:curvature}, we plot the magnitude of the dimensionless curvature of the material centerline, $\kappa=kD = \| \frac{\mathrm{d}^2\mathbf{r}(s)}{\mathrm{d}s^2} \| D$, as a function of the dimensionless arc-length coordinate, $s=S/D$. Again, a striking agreement is found between the experiments (dotted lines) and FEM (solid lines), in both the frictionless (Fig.~\ref{fig:curvature}(a)) and frictional cases (Fig.~\ref{fig:curvature}(b)). In Fig.~\ref{fig:curvature}(a), we also overlay the profile of the centerline curvature for the frictionless case (dashed line) predicted theoretically by the Kirchhoff rod model in Ref.~\cite{Audoly2007}. Surprisingly, the theoretical prediction yields a result that describes the experimental and simulation data remarkably well, albeit with small quantitative differences (the averaged difference in the curvature between the simulation data and the prediction is $\simeq 10\,\%$). Again, these deviations are expected, given the full 3D nature of our problem couple to the fact that the theory in Ref.~\cite{Audoly2007} was developed for loose knots, whereas we are considering tight configurations. 

\begin{figure}
    \centering
    \includegraphics[width=0.9\columnwidth]{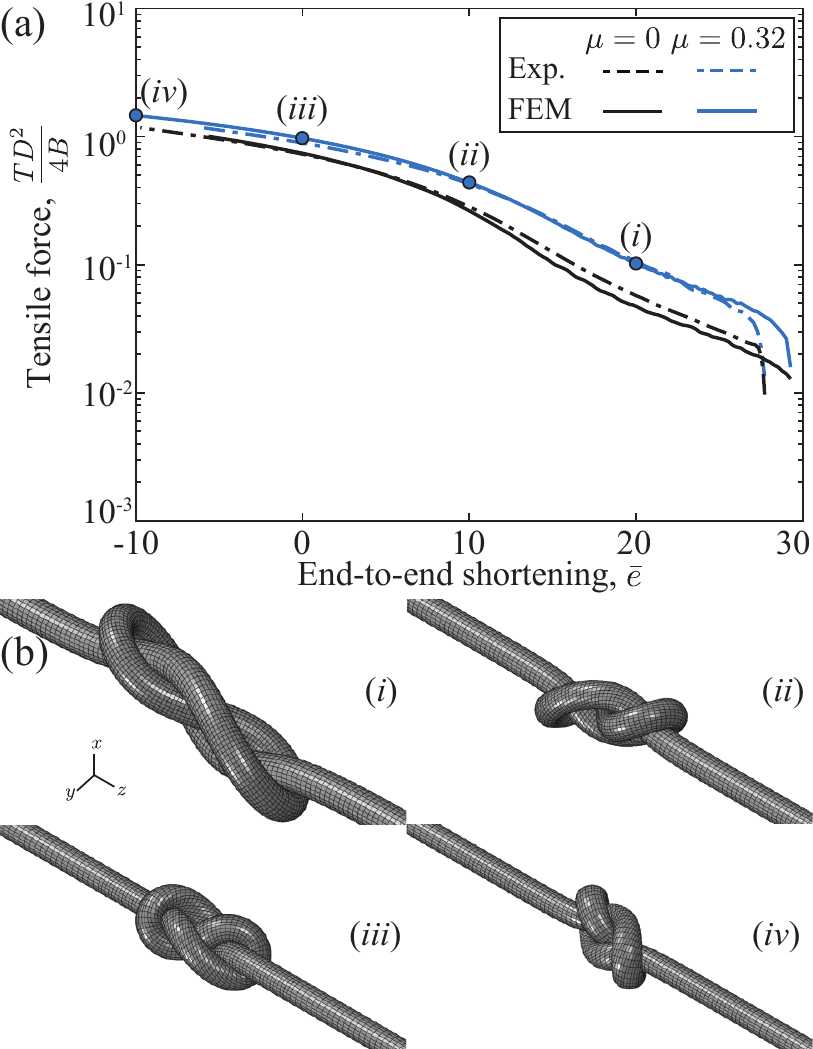}
    \caption{FEM and experimental results for the tying of a figure-of-eight knot on a rod of initial length $L_0=350\,\text{mm}$, and initial diameter $D=8.3\,\text{mm}$. (a) Tensile force versus end-to-end shortening of figure-of-eight knots (frictionless and frictional case). (b) Snapshots of the configurations of a computed figure-of-eight knot for consecutive values of the normalized end-to-end shortening, $\bar{e}=\{20, 10, 0, -10\}$. The corresponding data points, labeled as (\textit{i})-(\textit{iv}), are indicated in (a).}
    \label{fig:tensileforce_fig8}
\end{figure}

As another example to demonstrate the broader applicability of our protocol, in Fig.~\ref{fig:tensileforce_fig8}, we present a comparison for a different topology; a \textit{figure-of-eight knot}, with friction coefficients $\mu=\{0,0.32\}$. The rest configuration of the rod is $L_0=350\,\text{mm}$ and $D=8.3\,\text{mm}$. In Fig.~\ref{fig:tensileforce_fig8}(a), we provide the mechanical response of the figure-of-eight knot by plotting the dimensionless tensile force as a function of the dimensionless end-to-end shortening, whose range for the FEM simulations is within $-10 \le \bar{e} \le 29.5$. The corresponding experimental results for the same parameters are also included. In Fig.~\ref{fig:tensileforce_fig8}(b), we present four representative configurations of the FEM knot for $\mu=0.32$ and $\bar{e}=\{20, 10, 0, -10\}$. The data points corresponding to each configuration in Fig.~\ref{fig:tensileforce_fig8}(b) are located on the loading curve in Fig.~\ref{fig:tensileforce_fig8}(a). Again, the agreement between the FEM and the experiment is remarkable for both the frictionless and the frictional cases, further confirming the validity and high-fidelity of our FEM knot-tying approach.

\section{Conclusion}
\label{sec:conclusion}

We have presented a 3D FEM procedure to investigate tight elastic knots, both in the frictionless and frictional cases. As illustrative examples, we focused on the tight configurations of a simple overhand knot and a figure-of-eight knot. Our numerical results were found to be in excellent agreement with precision model experiments while showing deviations from an existing 1D theory for loose elastic knots~\cite{Audoly2007}. Our experimentally validated computational framework could be leveraged in the future to investigate quantities of tight elastic knots that are not readily available through experiments; \textit{e.g.}, the shape of the contact region, and cross-sectional deformation. Moreover, the versatility of the numerics should enable research efforts on knotted filaments with various physical ingredients; \textit{e.g.}, different constitutive descriptions, and friction models. We hope that the numerical framework that we have introduced will open new opportunities for more in-depth investigations of the mechanics of tight elastic knots.

\vspace{.5cm}

\begin{acknowledgment}
The authors are grateful to S. Neukirch and J.H. Maddocks for fruitful discussions.
P.J. acknowledges the support from Fonds National de la Recherche, Luxembourg (12439430). T.G.S. acknowledges financial support through a Grants-in-Aid for Overseas Research Fellowship from the Japan Society for the Promotion of Science (2019-60059).
\end{acknowledgment}

\bibliographystyle{asmems4}
\bibliography{asme2e}
\end{document}